\documentclass[twocolumn, prl, superscriptaddress]{revtex4}
\usepackage{amsmath}
\usepackage{subfigure}
\usepackage{natbib}
\usepackage{todonotes}
\usepackage{graphicx}
\usepackage[a4paper, total={7in, 10in}]{geometry}
\usepackage{hyperref}
\usepackage{multirow} 
\usepackage{footnote}

\hypersetup{
    colorlinks=true,
    linkcolor=black,
    filecolor=magenta,      
    urlcolor=blue,
    pdftitle={Overleaf Example},
    pdfpagemode=FullScreen,
    }

\usepackage{dcolumn}
\usepackage{bm}
\usepackage{orcidlink}
\usepackage{hyperref}
\usepackage{xcolor}
\usepackage{soul}
\sethlcolor{yellow}

\begin{document}

\title{Gravitational Wave Burst from Bremsstrahlung in Milky Way Can Discover Sub-Solar Dark Matter in Near Future}
\author{Samsuzzaman Afroz \orcidlink{0009-0004-4459-2981}}\email{samsuzzaman.afroz@tifr.res.in}
\author{Suvodip Mukherjee \orcidlink{0000-0002-3373-5236}}\email{suvodip@tifr.res.in}
\affiliation{Department of Astronomy and Astrophysics, Tata Institute of Fundamental Research, Mumbai 400005, India}

\begin{abstract}
What is Dark Matter, and what is its concentration in the Milky Way remains an open question in physics. We show that if a significant fraction of dark matter is composed of sub-solar mass primordial black holes (PBHs),  gravitational bremsstrahlung resulting from hyperbolic encounters between unbound PBHs within the galactic halos can generate distinctive chromatic gravitational-wave (GW) emission concentrated around the galactic dark matter halo, and it provides a direct window to discover such compact objects. We find that for both generalized NFW  and Einasto dark matter profiles of Milky Way, the signal-to-noise ratio can be more than five in one year of observation for the upcoming ground based GW observatories Cosmic Explorer if PBH dark matter fraction \( f_{\rm PBH} = 1 \) over the mass range \( 10^{-14} M_\odot \lesssim m_{\rm PBH} \lesssim 10^{-8} M_\odot \). Our results show that the Galactic Center could appear as a GW-bright source, enabling new insights into dark matter and its distribution. 
\end{abstract}

\maketitle

\textbf{\textit{Introduction:}} The identity of dark matter remains a fundamental mystery in modern physics \citep{Feng:2010gw,Zurek:2013wia,Cirelli:2024ssz,Arbey:2021gdg,Carr:2021bzv}. While the standard model of cosmology requires dark matter to account for nearly 27\% of the Universe’s energy budget, its underlying composition remains unknown \citep{Planck:2018vyg}. Among the numerous proposed candidates, primordial black holes (PBH) have emerged as a compelling possibility. Originally theorized in the 1970s as remnants from density fluctuations in the early Universe, PBH could have formed across a wide range of masses, from sub-planetary scales to many times the mass of the Sun \citep{Hawking:1975vcx, 1975ApJ...201....1C, Green:2014faa, Escriva:2022duf, Khlopov:1999ys, Green:2024bam}. Unlike particle-based dark matter candidates, PBH interact gravitationally, and if they are sufficiently abundant, they may constitute a significant or even dominant fraction of dark matter \citep{Carr:2020gox,Clesse:2016vqa,Carr:2020xqk,Green:2020jor}. In recent years, renewed interest in PBH has been driven by multiple observational and theoretical developments. Gravitational wave (GW) detections by the LIGO \citep{LIGOScientific:2016dsl}-Virgo \citep{VIRGO:2014yos}-KAGRA \citep{KAGRA:2020tym} (LVK) collaboration have revealed a population of binary black hole mergers whose masses and spins could be consistent with a primordial origin \citep{Sasaki:2016jop, Bird:2016dcv,Afroz:2024fzp,Afroz:2025efn}. Meanwhile, persistent null results in searches for weakly interacting massive particles (WIMPs) have increased the appeal of non particle dark matter candidates. Constraints on PBH from microlensing, cosmic microwave background anisotropies, and wide binary disruption still allow for viable PBH populations in certain mass windows particularly in the sub lunar to asteroid mass range, where direct detection remains challenging\citep{Arcadi:2017kky,Niikura:2017zjd,Ali-Haimoud:2016mbv,2014ApJ...790..159M,Carr:2016drx}. 

PBH, if present in sufficient numbers, would naturally reside within the dark matter halos of galaxies \citep{Afshordi:2003zb,Chisholm:2005vm,Carr:2018rid}. In these environments, mutual gravitational interactions between PBH can lead to occasional close hyperbolic encounters, during which a significant amount of energy is radiated as a GW burst by gravitational bremsstrahlung effect \citep{Kovacs:1978eu,Capozziello:2008ra,Steane:2023gme,Peters:1964zz,Hansen:1972jt,Peters:1970mx}. These bursts are short lived, highly eccentric, and non-periodic, with most of the energy released during the brief interval near periastron. Although individual bursts may be weak or rare, their cumulative contribution especially from high-density regions like the Galactic Center can give rise to a stochastic background or even resolvable signals within the sensitivity range of next generation detectors \citep{2017PhPl...24b2503H,Garcia-Bellido:2021jlq,Kocsis:2006hq,Cui:2021hlu}.

In this work, we focus on quantifying the GW brightness of the Milky Way arising from close encounters between mini black holes, under the assumption that they form a significant fraction of dark matter. The central question we ask is: How bright is the Galactic Center in gravitational waves, under various dark matter halo models and PBH scenarios?

To address this, we model the Galactic dark matter halo using a physically motivated density profiles, including the generalized Navarro-Frenk-White (gNFW) model, which allows for inner slope variations relevant to baryonic feedback and core formation \citep{Navarro:2008kc,Navarro:1996gj}. In addition, we incorporate the Einasto profile, which provides a good empirical fit to dark matter halos in cosmological simulations, particularly in the inner regions \citep{1965TrAlm...5...87E, Merritt:2005xc, 2024ApJ...962...81B,2010MNRAS.401.2318B}. We assume that PBH trace the dark matter distribution and compute the number density, encounter rate, and relative velocities of PBH in radial shells extending from the inner parsec out to 2 kiloparsecs (kpc). Importantly, we consider power-law model eccentricity distributions, which determine the spectral shape of the GW signal emitted during each flyby. This allows us to realistically capture the variation in GW emission across the Galactic halo.

Our findings show that under realistic conditions such as moderate clustering and sub-lunar PBH masses the inner Milky Way could produce a detectable signal which can be detected by Cosmic Explorer (CE) \citep{Reitze:2019iox}. This opens the possibility of using GW observations not only to constrain the PBH dark matter fraction, but also to probe the small scale structure of the Galactic halo, potentially offering new insights into the distribution and nature of dark matter in the central regions of galaxies.

\begin{figure*}[ht]
\centering
\includegraphics[height=4.5cm, width = 0.90\textwidth]{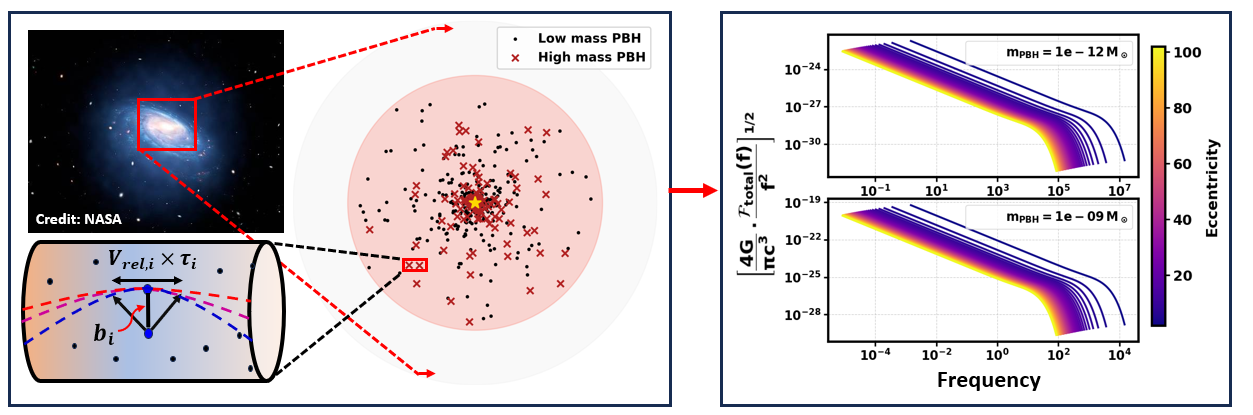}
\caption{\textbf{\textit{Gravitational-wave signal from PBH hyperbolic encounters in the Galactic halo.}} The left panel schematically illustrates the population of low-mass (black dots) and high-mass (red crosses) PBH distributed in the inner 2\,kpc of the Milky Way. A zoom-in to a representative region highlights the geometry of PBH\text{-}PBH hyperbolic encounters. The cylindrical inset shows a typical encounter: a PBH traverses a volume of radius \( b_i \) and length \( v_{\text{rel},i} \tau_i \), interacting with neighboring PBH (black dots). The dashed lines indicate different hyperbolic trajectories (corresponding to different values of eccentricity), with emission concentrated near periapsis over a timescale \( \tau_i \). The right panel presents the frequency-dependent quantity $\left[(4G/\pi c^3) \cdot \mathcal{F}_{\mathrm{total}}(f)/f^2 \right]^{1/2}$, which characterizes the GW signal strength, computed over one year of cumulative emission assuming $f_{\rm PBH} = 1$, for two representative PBH masses: \( 10^{-12}\,M_\odot \) (top) and \( 10^{-9}\,M_\odot \) (bottom). Signal curves are color-coded by orbital eccentricity, with higher \( e \) (lighter shades) enhancing low-frequency emission. The dark matter density is modeled using a generalized NFW profile, and the flux is integrated over spherical shells. Together, this figure illustrates how small-scale PBH dynamics in the Galactic Center can imprint detectable GW signatures across a wide frequency range.}
\label{fig:Schematic}
\end{figure*}

\vspace{0.5em}
\noindent
\textbf{\textit{Gravitational Wave Emission from single Hyperbolic Encounters:}} GW bursts can be produced during close hyperbolic encounters between compact objects, such as PBH, neutron stars, or black holes. These unbound interactions occur in dark matter halos or dense stellar environments, where objects approach one another with high relative velocities and emit gravitational radiation near the point of closest approach.

For an encounter between two compact objects with relative velocity \( v_{\text{rel}} \), the closest distance or periapsis for an orbit with eccentricity \( e > 1 \) is given by \cite{1977ApJ...216..610T}
\begin{equation}
r_{\min} = \frac{G m_{tot}}{v_{\text{rel}}^2}(e - 1),
\end{equation}
where \( m_{tot} = m_1 + m_2 \) is the total mass of the system. The characteristic duration of the encounter, over which most GW emission occurs, is approximated by \cite{1977ApJ...216..610T}
\begin{equation}\label{eq:tau}
\tau = \frac{r_{\min}^{3/2}}{\sqrt{G m_{tot} (1 + e)}}.
\end{equation}

The characteristic duration of the encounter \( \tau \) sets the peak GW frequency as \( f_{\rm peak} \sim 1/\tau \). For fixed velocity, more massive binaries have longer \( \tau \), emitting at lower frequencies, while higher eccentricities or lighter masses lead to shorter \( \tau \) and higher-frequency bursts. Thus, PBH encounters span a broad frequency range, from mHz to kHz, depending on mass and velocity. We compute the spectral energy emitted per unit frequency using the Peters-Mathews formalism \cite{1977ApJ...216..610T}. Defining the dimensionless variable \( u = 2\pi f \tau \), the frequency structure of the burst is captured by Bessel function combinations:
\begin{equation}
\begin{split}
t_1 &= u^2 K_2(u) - u K_1(u), \\
t_2 &= 2 u^2 K_1(u) + u K_0(u), \\
t_3 &= u K_0(u),
\end{split}
\end{equation}
where \( K_n \) are modified Bessel functions of the second kind. The resulting GW spectral luminosity is given by \cite{1977ApJ...216..610T}


\begin{align}
\begin{split}
    \frac{dL}{df} &\equiv \frac{1}{\tau} \times \frac{du}{df} \times \frac{dE}{du}, \\ &= \frac{2\pi G^{7/2} \sqrt{m_{\rm tot}} (m_1 m_2)^2 \sqrt{e}}{c^5 r_{\min}^{7/2}} \times \frac{8}{15\pi} \left(12 t_1^2 + 3 t_2^2 + t_3^2 \right).
\end{split}
\end{align}

To convert this to the observed energy flux at the detector, one includes the geometric dilution factor associated with the source-observer separation \citep{DeVittori:2012da,Capozziello:2008ra}. Assuming the observer is at distance \( d \), the flux is given by:
\begin{equation}
\mathcal{F}(f) = \frac{1}{4\pi d^2} \times \frac{dL}{df}.
\end{equation}

The actual flux received by a detector depends on the relative orientation between the orbital plane and the line of sight. The GW emission from hyperbolic encounters is anisotropic, peaking in the direction perpendicular to the orbital plane. To account for this, one often considers the angle-averaged flux, obtained by integrating over all possible orientations of the binary system relative to the observer. This averaging is especially relevant when dealing with unresolved or randomly oriented sources, as in cosmological populations of PBH. In the case of isotropic distributions, the angle-averaged flux provides a robust estimator for computing the stochastic GW background or expected burst rates.

\vspace{0.5em}
\noindent
\textbf{\textit{Eccentricity Modelling for Hyperbolic Encounters:}} In scenarios involving unbound compact object interactions-such as PBH flybys or dynamical GW bursts-the binaries follow hyperbolic trajectories characterized by eccentricities \( e > 1 \). The distribution of eccentricity in such encounters encodes critical information about the impact parameters, relative velocities, and gravitational focusing of the interacting objects \citep{Merino:2009yj, Kocsis:2011jy}. For a given initial condition on the initial velocity, impact factor of an object the eccentricity can be calculated. However, since these quantities are unknown and an ensemble of particles are going to follow a distribution, we consider a distribution of eccentricity of the particles in the remaining work. 

In this work, we adopt a power-law model for the eccentricity distribution, motivated by gravitational scattering theory, where larger impact parameters (and hence larger values of \( e \)) are statistically suppressed \citep{1983ApJ...268..319H, Samsing:2017xmd}. The probability distribution takes the form:
\[
P(e) \propto (e - 1)^{-\gamma}, \quad e \in [e_{\min}, e_{\max}),
\]
where \( \gamma > 0 \) controls the steepness of the distribution, and the cutoff parameters \( e_{\min} \) and \( e_{\max} \) define the physically motivated range of eccentricities. 

In our analysis, we fix \( \gamma = 3 \), corresponding to a moderate suppression of high-eccentricity events while retaining significant contributions from near-parabolic encounters. To assess the robustness of our results, we also show representative cases with \( \gamma = 2 \) and \( \gamma = 4 \), demonstrating how variations in the eccentricity distribution affect the signal. We set \( e_{\max} = 1000 \) and vary \( e_{\min} \) to explore its impact on the GW burst detectability.

\vspace{0.5em}
\noindent
\textbf{\textit{Dark Matter Halo Model of Milky Way:}} Several density profiles have been proposed to model the distribution of DM in galactic halos. Among these, we adopt the gNFW and Einasto profiles in this work. These models are well motivated by numerical simulations and have been shown to provide good fits to the Milky Way’s rotation curve \citep{Pato:2015dua,Li:2025mqx,2021A&A...654A..25J,Navarro:1995iw}.

The gNFW profile introduces an inner slope parameter \( \beta \), allowing for more flexibility in describing the inner halo structure \citep{Navarro:2008kc,Navarro:1996gj}. Its density profile is given by
\begin{equation}
\rho_{\rm gNFW}(r) = \frac{M_0}{4\pi\,r_s^3}\,\frac{1}{(r/r_s)^\beta\,(1 + r/r_s)^{3-\beta}},
\end{equation}
where \( M_0 \) is the normalization mass, \( r_s \) is the scale radius, and \( \beta \) determines the inner slope. This form accommodates both cored (\( \beta \to 0 \)) and cuspy profiles.

The Einasto profile, which features an exponential radial dependence, avoids central cusps and better matches DM halo profiles found in simulations \citep{1965TrAlm...5...87E, Merritt:2005xc}
\begin{equation}
\rho_{\rm Ein}(r) = \frac{M_0}{4\pi\,r_s^3}\,\exp\left[-\left(\frac{r}{r_s}\right)^\alpha\right],
\end{equation}
where \( \alpha \) controls the shape of the profile's fall-off.

We model the Milky Way DM halo using the gNFW and Einasto profiles, with best-fit parameters from recent rotation-curve fits \citep{2024MNRAS.528..693O}. For gNFW, we use \( M_0 = (3.21 \pm 0.07) \times 10^{11}\, M_\odot \), \( r_s = 5.26^{+0.15}_{-0.11} \) kpc, and \( \beta = 0.0258^{+0.0416}_{-0.0192} \). For the Einasto profile, the parameters are \( M_0 = (0.62^{+0.12}_{-0.11}) \times 10^{11}\, M_\odot \), \( r_s = 3.86^{+0.35}_{-0.38} \) kpc, and \( \alpha = 0.91^{+0.04}_{-0.05} \). These values capture the best estimates for the DM distribution in the Milky Way and serve as the baseline in our subsequent analysis.

\vspace{0.5em}
\noindent
\textbf{\textit{Gravitational-Wave Flux from PBH Interactions in a Dark Matter Halo:}} To compute the total GW emission from PBH-PBH encounters within a dark matter halo, we divide the spatial domain from \( r = 0 \) to \( r = 2\,\mathrm{kpc} \) which encompasses the region contributing most significantly to the SNR, beyond which the signal becomes increasingly suppressed due to lower encounter rates at larger distances-into a series of concentric spherical shells. These shells are logarithmically spaced to ensure higher resolution in the inner halo, where the density gradients are steepest. The \( i \)-th shell is bounded by inner and outer radious \( r_{i} \) and \( r_{i+1} \), with midpoint  \( r_{\text{mid, i}} = (r_i + r_{i+1})/2 \).

The volume of the \( i \)-th shell is given by
\begin{equation}
V_{\text{shell, i}} = \frac{4\pi}{3} (r_{i+1}^3 - r_i^3),
\end{equation}
and the mass of dark matter contained within it is calculated as
\begin{equation}
M_{\text{shell,i}} = M_{\text{DM}}(\le r_{i+1}) - M_{\text{DM}}(\le r_i),
\end{equation}
where \( M_{\text{DM}}(\le r) \) is the enclosed mass up to radius \( r \), obtained by numerically integrating the DM density profile
\begin{equation}
M_{\text{DM}}(\le r) = \int_0^r 4\pi r'^2 \rho(r') \, dr'.
\end{equation}
Assuming all dark matter is composed of PBH of mass \( m_1 \), the number of PBH in a shell is
\begin{equation}
N_{\text{PBH,i}} = \frac{M_{\text{shell,i}}}{m_1},
\end{equation}
and the corresponding number density is
\begin{equation}
n_{\text{PBH,i}} = \frac{N_{\text{PBH,i}}}{V_{\text{shell,i}}}.
\end{equation}

The typical relative velocity (i.e., the local velocity dispersion) between PBH at radius $r_{\text{mid,i}}$ is estimated via the virial relation,
\begin{equation}
v_{\text{rel,i}} = \sqrt{ \frac{G M(\le r_{\text{mid,i}})}{r_{\text{mid,i}}} }.
\end{equation}
This velocity dispersion characterizes the distribution of relative velocities in the halo and governs the encounter dynamics. The impact parameter is taken as the average inter-PBH separation, \( b_i = n_{\text{PBH,i}}^{-1/3} \), which sets the gravitational focusing scale. The corresponding cross-section for encounters is
\begin{equation}
\sigma_{\text{enc,i}} = \pi b_i^2.
\end{equation}

Using these quantities, we can calculate the total number of PBH-PBH encounters on an average over a time window $\tau$, which is the typical timescale of the emission of the GW bremsstrahlung signal (see Eq. \eqref{eq:tau}), and the objects to get deflected. The total number of such encounters within a shell (denoted by index 'i')  over the observation time \( T \) is then computed as
\begin{equation}
\begin{split}
N_{\text{enc,i}} = 
\underbrace{N_{\text{PBH,i}}}_{\text{Total PBH}} \times 
\underbrace{ \left( n_{\text{PBH,i}} v_{\text{rel,i}} \sigma_{\text{enc,i}} \tau_i \right) }_{\substack{\text{Encounter rate per PBH} \\ \text{per time window } \tau_i}} \times 
\underbrace{ \left( \frac{T}{\tau_i} \right) }_{\substack{\text{Number of} \\ \text{time windows}}}.
\end{split}
\end{equation}
This expression shows that the total number of encounters is the product of three physical ingredients: (1) the number of PBH available to participate in encounters, (2) the rate at which an individual PBH undergoes interactions over a time interval \( \tau \), and (3) the number of such intervals within the total observation time \( T \).

Each of these encounters emits a GW burst, characterized by the spectral luminosity \( dL/df \) as described previously. To compute the observed flux, we account for the geometric dilution of the signal from each shell as seen by an observer located at \( r_{\text{obs}} = 8.23\,\mathrm{kpc} \) from the Galactic center \citep{2023MNRAS.519..948L}. Since each spherical shell contributes GW signals from both the near and far sides with respect to the observer, we use the angular-averaged inverse-square distance factor,
\begin{equation}
\left\langle \frac{1}{d^2} \right\rangle = \frac{1}{2 r_{\text{mid}} r_{\text{obs}}} \log \left( \frac{r_{\text{mid}} + r_{\text{obs}}}{|r_{\text{mid}} - r_{\text{obs}}|} \right),
\end{equation}
which correctly averages the flux contribution over all angles between the shell and the observer. This ensures that both the near-side and far-side contributions from each shell are included consistently in the final flux calculation. The total energy flux at frequency \( f \) from the shell is
\begin{equation}
\mathcal{F}_i(f) = \left\langle \frac{1}{d^2} \right\rangle_i \times N_{\text{enc},i} \times \left( \frac{dL}{df} \right)_i.
\end{equation}
Finally, summing over all radial shells between 0 and 2 kpc gives the full GW flux spectrum:
\begin{equation}
\mathcal{F}_{\text{total}}(f) = \sum_i \mathcal{F}_i(f) = \sum_i N_{\text{enc},i} \times \left( \frac{dL}{df} \right)_i \times \left\langle \frac{1}{d^2} \right\rangle_i.
\end{equation}

\vspace{0.5em}
\noindent
\textbf{\textit{Detectability: Signal-to-Noise Ratio:}} To assess the detectability of the GW background generated by PBH-PBH encounters in a dark matter halo, we compute the SNR of detecting the total flux as \citep{Allen:1997ad,Sutton:2013ooa}
\begin{equation}
\text{SNR}^2 = 
\frac{16G}{\pi c^3}\int_{f_{\min}}^{f_{\max}} \frac{{ \mathcal{F}_{\text{total}}(f)}{}}{f^3 S_n(f)} \, df,
\end{equation}
where the limits \( f_{\min} \) and \( f_{\max} \) are determined by the intersection of the signal bandwidth with the detector's sensitivity band having a noise power spectral density \( S_n(f) \).

\begin{figure*}[ht]
\centering
\includegraphics[height=5.5cm, width = 0.48\textwidth]{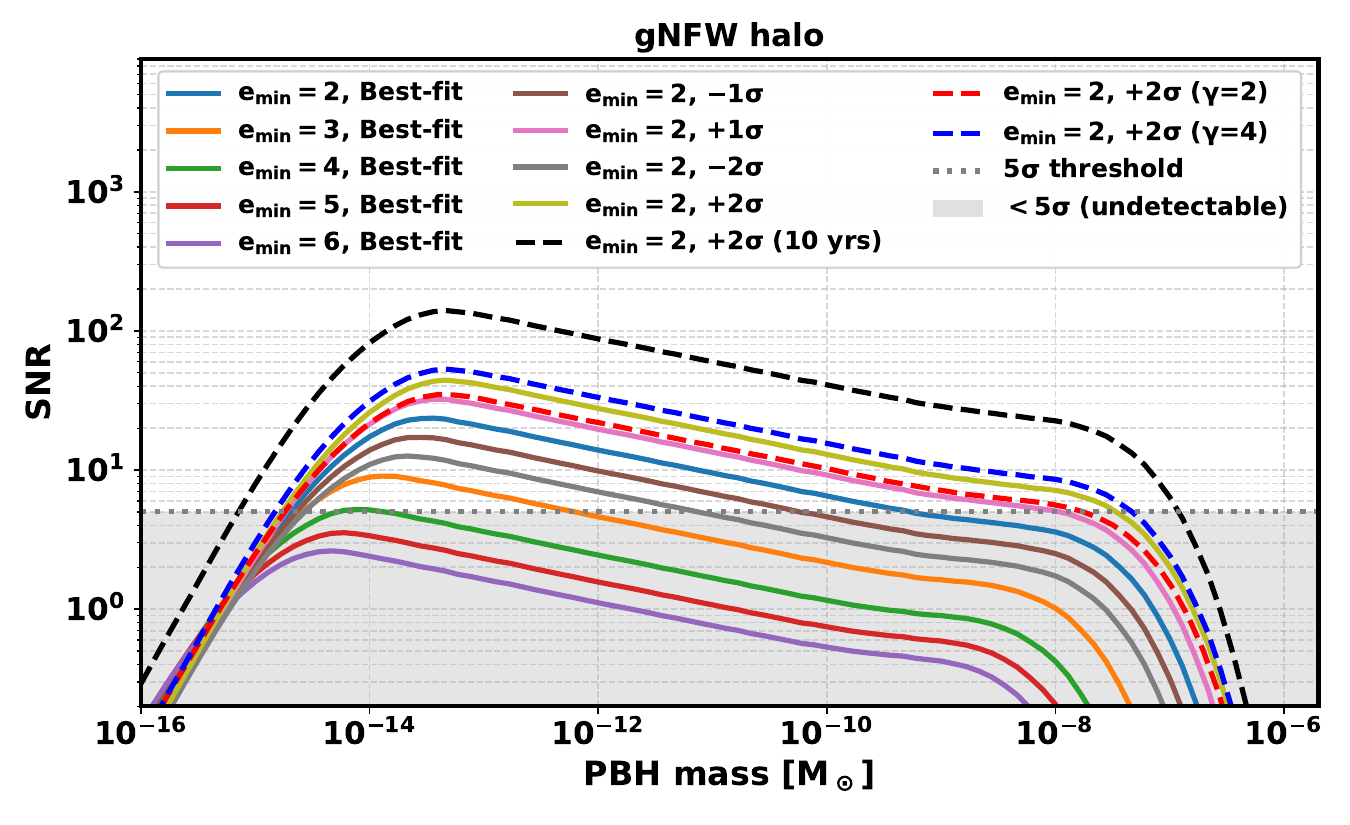}
\includegraphics[height=5.5cm, width = 0.48\textwidth]{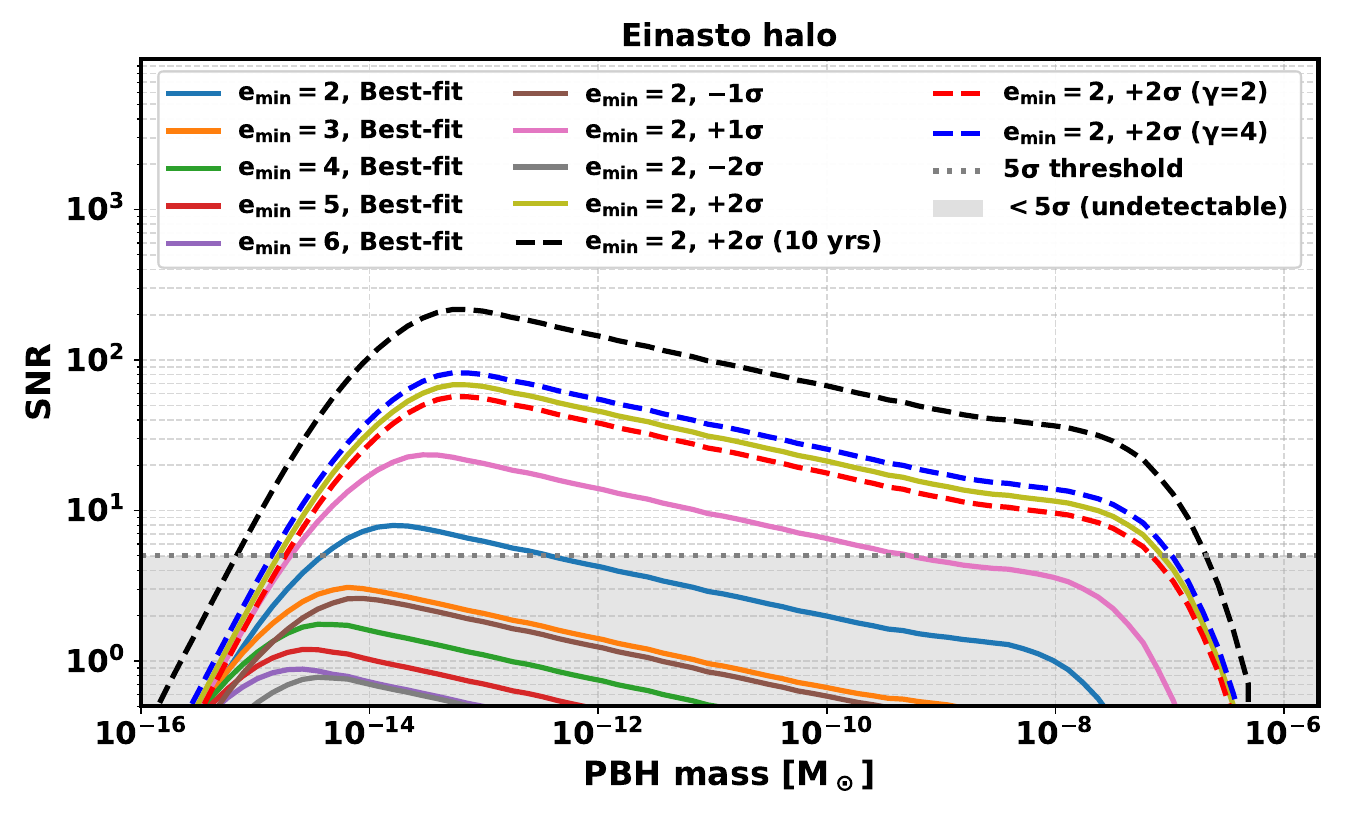}
\caption{Signal-to-noise ratio as a function of PBH mass for GW bursts from hyperbolic encounters of PBH, shown for the CE under two different dark matter halo models: gNFW (left) and Einasto (right). In each panel, the SNR is computed assuming a PBH abundance of \( f_{\rm PBH} = 1 \). We show results for the best-fit halo profiles across a range of minimum encounter eccentricities (\( e_{\min} = 2 \) to 6), where lower \( e_{\min} \) values lead to higher SNR. To illustrate uncertainties in halo modeling, we also include curves corresponding to \( \pm1\sigma \) and \( \pm2\sigma \) deviations at fixed \( e_{\min} = 2 \). The black dashed lines represent the impact of an extended observation time of 10 years, shown for the \( +2\sigma \), \( e_{\min} = 2 \) scenario. Additionally, we explore the effect of varying the power-law slope of the eccentricity distribution (\( \gamma \)), with results for \( \gamma = 2 \) and \( \gamma = 4 \) shown alongside the default case \( \gamma = 3 \). Unless otherwise stated, all curves assume \( \gamma = 3 \) and a 1-year observation time.}
\label{fig:SNR_multi_panel}
\end{figure*}

\vspace{0.5em}
\noindent
\textbf{\textit{Results:}} In this study, we present GW signal predictions from hyperbolic encounters between PBH distributed within the inner 2~kpc of the Milky Way's dark matter halo. We consider two canonical dark matter profiles-the gNFW and the Einasto profile-and compute the expected GW strain spectra assuming that PBH constitute the entirety of the dark matter content (\( f_{\rm PBH} = 1 \)). The encounter rates are calculated shell-by-shell based on the local PBH density, relative velocity, and impact parameters, and the resulting GW flux is used to derive the strain-related quantity \( \left[(4G/\pi c^3) \cdot \mathcal{F}_{\mathrm{total}}(f)/f^2 \right]^{1/2} \) observed over a 1-year integration period. Right side of Figure~\ref{fig:Schematic} shows this frequency-dependent signal \( \left[(4G/\pi c^3) \cdot \mathcal{F}_{\mathrm{total}}(f)/f^2 \right]^{1/2} \) arising from hyperbolic encounters of PBH, for a range of orbital eccentricities at two fixed PBH masses. The top and bottom panels correspond to \( m_{\rm PBH} = 10^{-12}\,M_\odot \) and \( 10^{-9}\,M_\odot \), respectively. For each mass, we compute the strain for fifty representative eccentricity values ranging from \( e = 2 \) to \( 102 \), as indicated by the color gradient in each panel. The figure illustrates how eccentricity strongly impacts both the amplitude and frequency of the resulting GW signals. Lower eccentricities lead to more pronounced and high-frequency strain, reflecting longer interaction times and greater energy release during close approaches. This effect is more prominent for higher-mass PBH, which also yield overall stronger signals due to their larger gravitational influence. In contrast, at lower masses, the signal shifts to higher frequencies and becomes weaker, consistent with more rapid flybys and shorter interaction timescales.

In Figure~\ref{fig:SNR_multi_panel}, we present the SNR as a function of PBH mass for CE \citep{Reitze:2019iox}, which is optimized for high-frequency bursts in the \( \sim 5\,\mathrm{Hz} \)–\( 5\,\mathrm{kHz} \) band, relevant for lighter PBHs or faster flybys. We consider both gNFW (left) and Einasto (right) dark matter halo profiles, including best-fit parameters as well as their \( \pm 1\sigma \) and \( \pm 2\sigma \) variations. The SNR is computed assuming a fiducial observation time of \( T_{\rm obs} = 1\,\mathrm{yr} \) and a PBH dark matter fraction \( f_{\rm PBH} = 1 \). We vary the minimum eccentricity parameter \( e_{\min} \) in the distribution \( P(e) \propto (e - 1)^{-\gamma} \), fixing \( \gamma = 3 \), and show results for \( e_{\min} = 2, 3, 4, 5, 6 \). To assess halo modeling uncertainties, we also show SNR curves for \( e_{\min} = 2 \) with profile parameters varied within \( \pm 1\sigma \) and \( \pm 2\sigma \) ranges. Finally, to explore the impact of the eccentricity slope, we include cases with \( \gamma = 2 \) and \( \gamma = 4 \).

\vspace{0.5em}
\noindent
\textit{Mass dependence:} The SNR as a function of PBH mass exhibits a non-monotonic behavior. At very low masses, the SNR is suppressed due to high-frequency GW bursts falling outside CE's sensitive frequency band. As mass increases, the burst frequency enters the detector's optimal range and the GW flux per encounter increases, leading to a rise in SNR. The SNR reaches a maximum around $m \sim 10^{-14} M_\odot$, where the trade-off between flux strength and the number of encounters is optimized. Beyond the peak, the SNR decreases with mass as the encounter rate drops faster than the flux gain per event. While the GW flux per encounter grows linearly with mass, the number of encounters scales as $m^{-4/3}$, leading to $\text{SNR}^2 \propto m^{-1/3}$. This inverse scaling dominates at higher masses, producing a bell-shaped SNR curve on a log-log plot.

\vspace{0.5em}
\noindent
\textit{Eccentricity distribution:} Lowering \( e_{\min} \) increases the SNR across all masses, as near-parabolic encounters (with smaller periapsis distances) produce more efficient GW emission. In contrast, increasing \( e_{\min} \) suppresses these strong encounters, reducing detectability. While we fix the power-law slope at \( \gamma = 3 \) in the default scenario, we also present results for \( \gamma = 2 \) and \( \gamma = 4 \) to illustrate how the SNR depends on the eccentricity distribution: smaller \( \gamma \) increases the weight of high-eccentricity events and lowers the SNR, whereas larger \( \gamma \) suppresses such events and slightly boosts the SNR (see Fig.~\ref{fig:SNR_multi_panel}).

\vspace{0.5em}
\noindent
\textit{Dark matter profile uncertainties:} Variations in the dark matter halo profile lead to significant changes in the local density, and hence the encounter rate. These variations translate to up to a factor of \( \sim 2{-}3 \) difference in the SNR. This highlights the importance of accurately modeling the inner halo structure when interpreting burst signals.

The SNR scales as \( \mathrm{SNR} \propto f_{\rm PBH} \sqrt{T_{\rm obs}} \), since the number of encounters grows as \( f_{\rm PBH}^2 T_{\rm obs} \). Thus, lower PBH fractions or shorter runs reduce detectability, while longer observations can enhance sensitivity to subdominant PBH populations.

\vspace{0.5em}
\noindent

\textbf{\textit{Conclusions:}} Our analysis demonstrates that hyperbolic GW bursts from unbound PBH interactions in the Milky Way can produce detectable signals from the upcoming ground-based GW detectors across a broad PBH mass range, even under conservative assumptions. The strain spectrum and resulting SNR depend sensitively on the PBH mass, the eccentricity distribution of the encounters, and the DM halo profile. Steep inner density slopes, as seen in gNFW profiles, enhance the encounter rate and increase signal strength compared to cored profiles like Einasto in the best-fit case.

These results open a novel observational avenue to probe sub-lunar to asteroid-mass PBH within our Galaxy-mass ranges that are challenging to access with traditional cosmological probes such as CMB distortions, microlensing, or large-scale structure. Unlike those methods, which are largely sensitive to PBH over cosmological distances, the signals considered here arise from the local distribution of PBH, primarily within $\sim 2\,\mathrm{kpc}$ of the Galactic Center. This makes the resulting signal inherently anisotropic and offering a complementary and potentially more direct handle on Galactic PBH populations which is currently unknown.

Looking forward, targeted anisotropic searches focusing on the Galactic Center direction could significantly enhance sensitivity. The strongly peaked spatial distribution of these signals breaks the isotropy assumed in most stochastic background searches and may allow directional filtering techniques to isolate them from instrumental and astrophysical backgrounds. Moreover, any detection of such a signal would provide unique insight into the low-eccentricity tail of hyperbolic PBH binaries-an aspect that remains unconstrained observationally.

This work motivates the inclusion of hyperbolic burst templates in the data analysis pipelines, particularly those optimized for transient signals from the Galactic Center. A future detection or upper limit on such bursts would not only constrain the PBH abundance $f_{\rm PBH}$ but also shed light on the shape of the inner DM halo and the dynamics of compact objects in high-density regions of galaxies. By enabling Galactic-scale tests of dark matter microphysics, this approach could offer one of the few direct probes of unbound compact object interactions in the strong-field regime of gravity.

\textbf{\textit{Acknowledgments:}} This work is part of the \texttt{⟨data|theory⟩ Universe-Lab}, supported by TIFR and the Department of Atomic Energy, Government of India. The authors express gratitude to the system administrator of the computer cluster of \texttt{⟨data|theory⟩ Universe-Lab} for maintaining the computing resources. We acknowledge the use of the following packages in this work: Numpy \cite{harris2020array}, Scipy \cite{virtanen2020scipy},  and Matplotlib \cite{Hunter:2007}.

\bibliographystyle{unsrt}
\bibliography{references}
\end{document}